\documentclass[final,5p,twocolumn,authoryear]{elsarticle}
\usepackage{ifthen}
\newboolean{publ}
\setboolean{publ}{true}

\makeatletter
\def\ps@pprintTitle{%
  \let\@oddhead\@empty
  \let\@evenhead\@empty
  \def\@oddfoot{\footnotesize\itshape
       Published in: \ifx\@journal\@empty Elsevier
       \else\@journal\fi\hfill\today}%
  \let\@evenfoot\@oddfoot
}
\makeatother

\usepackage{setspace} 
\usepackage{amssymb}				
\usepackage{graphicx}			
\ifthenelse{\boolean{publ}}{}{%
	\usepackage{lineno}%
	\modulolinenumbers[5]%
}
\usepackage[colorlinks]{hyperref}	 
\usepackage{textcomp}			

\begin{document}

\title{An Ancestral Axial Twist Explains the Contralateral Forebrain and the Optic Chiasm in Vertebrates\tnoteref{t1}}

\tnotetext[t1]{Cite as: de Lussanet, MHE \& Osse, JWM (2012) An ancestral axial twist explains the contralateral forebrain and the optic chiasm in vertebrates.\ {\em Animal Biology} {\bf 62}(2):193-216.\ DOI: 10.1163/157075611X617102. \\This is the accepted, peer-reviewed version, without the Journal's formatting and the small editorial textual improvements. Some of the figures are color versions. \\Note that one correction was made in May 2014: see footnote \ref{fnoteCorr}.}

\author[wwu]{Marc H. E. de Lussanet\corref{cor1}} 
\ead{lussanet@wwu.de}
\author[wag]{Jan W. M. Osse} 

\journal{Animal Biology}

\cortext[cor1]{Corresponding author} 
\address[wwu]{Institute of Psychology, Westf. Wilhelms-Universit{\"a}t, Fliednerstra{\ss}e 21, 48149 M{\"u}nster, Germany} 
\address[wag]{Bennekomseweg 83, 6704 AH Wageningen, the Netherlands} 

\ifthenelse{\boolean{publ}}{}{\onehalfspacing}

\begin{abstract}
Among the best-known facts of the brain are the contralateral visual, auditory, sensational, and motor mappings in the forebrain. How and why did these evolve? The few theories to this question provide functional answers, such as better networks for visuomotor control. However, these theories contradict the data, as discussed here. 

Instead we propose that a 90-deg turn on the left side evolved in a common ancestor of all vertebrates. Compensatory migrations of the tissues during development restore body symmetry. Eyes, nostrils and forebrain compensate in the direction of the turn, whereas more caudal structures migrate in the opposite direction. As a result of these opposite migrations the forebrain becomes crossed and inverted with respect to the rest of the nervous system.

We show that such compensatory migratory movements can indeed be observed in the zebrafish (\emph{Danio rerio}) and the chick (\emph{Gallus gallus}). With a model we show how the axial twist hypothesis predicts that an optic chiasm should develop on the ventral side of the brain, whereas the olfactory tract should be uncrossed. 

In addition, the hypothesis explains the decussation of the trochlear nerve, why olfaction is non-crossed, why the cerebellar hemispheres represent the ipsilateral bodyside, why in sharks the forebrain halves each represent the ipsilateral eye, why the heart and other inner organs are asymmetric in the body. Due to the poor fossil record, the possible evolutionary scenarios remain speculative. Molecular evidence does support the hypothesis. The findings may shed new insight on the problematic structure of the forebrain.
\end{abstract}



\maketitle 
\ifthenelse{\boolean{publ}}{}{\linenumbers}

\section{Introduction}
\label{secIntro}

Contralateral means ``on the other side''. It refers to the fact that each side of the brain receives its input from the eye (or visual field) on the opposite side, and also controls and senses the opposite bodyside. The contralateral organisation of the forebrain is a general and prominent vertebrate feature, but explanations are rare and usually only address either the motor or the visual aspect \citep{vulliemoz2005.87-99,shinbrot2008.1278-1292}. Although much information has become available about molecular and genetic mechanisms, the general pattern remains puzzling \citep{stern2002.447-451, lee2004.947-960, jeffery2005.721-753}. The only theory to date that tries to explain the general pattern of contralateral representations in the brain, is by \citet{ramon-y-cajal1898.15-65}. Ram\'on y Cajal was the first to develop a conceptual framework to explain the pattern of contralateral representations and neuronal decussations\footnote{Decussation comes from the latin numeral X (ten), is used to refer to midline-crossings of axons or nerves. Chiasm is derived from the greek letter $\chi$, and is used for the midline crossing of the optic tract} in a functional manner \citep{ramon-y-cajal1898.15-65,llinas2003.77-80, vulliemoz2005.87-99, loosemore2009.375-382}. He carefully described the fibre crossings in several species of bony fish, amphibians, reptiles, birds and mammals. 

\begin{figure}[htb!]	 \centering
\includegraphics[width=7.5cm]{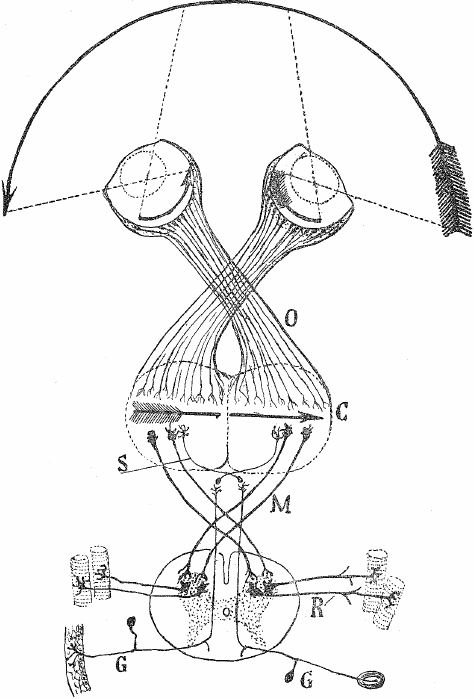}
\caption{Ram\'on y Cajal's theory for the contralateral organisation of the forebrain. He proposed that the contralateral somatosensory and motor representations are adaptations to the visual system (see text for explanation). O, optic tract; C, primary and secondary visual centres; M, decussating motor pathways; S, decussating sensory pathways; R, motor efferents of the spine; G, spinal ganglions and sensory afferents. Note that Cajal drew the spinal section upside down and the cortical hemispheres as fused. The secondary visual centres presumably refer to primary motor and somatosensory cortex. Reproduced from \citet{ramon-y-cajal1898.15-65}.}
\label{figCajal} \end{figure}

Cajal's theory can best be explained at hand of one of his own figures (reproduced in fig.\ \ref{figCajal}). Central to his theory is the hypothesis that the brain creates a panoramic representation of the outer world. In a lateral-eyed animal, the two retinas have non-overlapping projections of the world. The retina inverts the world. One way to create a complete, aligned projection from the two hemifields is to cross each optic tract to the contralateral side of the brain (fig.\ \ref{figCajal}, upper part). 

Animals with frontally positioned eyes have overlapping visual fields. In such animals, including humans, not all optic tract fibers cross the midline. Instead, each hemisphere of the brain represents the contralateral visual hemifield of both eyes \citep{jeffery2005.721-753}. It is generally accepted that this organisation is highly advantageous for depth perception. The local mismatch between the images on corresponding locations of the two retinas (known as disparity) provides accurate information about the distance of objects with respect to the eyes \citep{regan2000.book}. Already according to Cajal, the incomplete crossing of the retinal projections in the optic chiasm of frontal-eyed land vertebrates is a later adaptation to achieve depth vision from binocular disparity. 

To complete his theory, Cajal proposed that the contralateral motor and somatosensory representations in the forebrain are an adaptation to the contralateral visual representation (fig.\ \ref{figCajal}, lower part). He argued that the motor and somatosensory representations must be crossed in order to match the right visual hemifield with the right bodyside and the left visual hemifield with the left bodyside.

Cajal's theory is still popular and, at first sight, elegant. However, a closer look reveals several problems. It is questionable, in the first place, whether a continuous, aligned projection in the brain is an advantage at all. All vertebrates can move their eyes, and moving the eyes makes the retinal images shift with respect to each other, so a true alignment of the visual fields is impossible. 

A more serious problem of Cajal's theory is that the hemispheres of the forebrain are comparatively badly connected. If an aligned projection would have a special evolutionary advantage, one should have expected that the primary visual cortex on each side of the brain should be a most densely interconnected part of the brain, especially in lateral eyed species. There is no evidence for this. The corpus callosum, which Ram\'on y Cajal thought to fulfil this function, only evolved in mammals. In other vertebrates the cortical hemispheres are connected only by the relatively small anterior commissure.

At the level of the sensorimotor representations in the brain, Cajal's explanation is also seriously problematic. The theory assumes that it is optimal to connect each side of the visual world with the ipsilateral sensorimotor system. However, theoretical models show that even for the simplest imaginable animal it is optimal to have bilateral connections between visual and motor systems \citep{braitenberg1984.book}. \citet{flechsig1899.III-VI} already questioned this part of Cajal's theory in his foreword to the german translation of \citep{ramon-y-cajal1898.15-65}, arguing that there is no direct projection of the primary visual cortex to the motor cortex. Some studies have used similar reasoning as Ram\'on y Cajal, but found that the crossed connections of visual hemifield and body and leg musculature is optimal \citep{loeb1918.book,bertin1994.book}. 

Functional explanations focus on vertebrates. This is problematic, because molluscs (cephalopods) and arthropods also evolved highly sophisticated visual and motor systems. In spite of the well-developed interhemispheric connections, none of these clades seems to have evolved systematic contralateral representations \citep{budelmann1995.115-138,hausen1984.523-555}.

According to an entirely different approach, Ebbesson's parcellation theory, lateralisation is an epiphenomenon of an increase in brain size \citep{ebbesson1980.179-212}. The primitive condition, according to this view, is a bilateral representation, whereas the evolution of contralateral representations is coincidental and as likely as the evolution of ipsilateral representations \citep{ebbesson1980.483-495}. However, there are examples of extensive bilateral connections in the optic chiasm, in several lineages of vertebrates. Thus, the theory implies that a complete decussation in the optic chiasm of vertebrates should have evolved independently many times. It seems highly unlikely that this could have happened without altogether losing the optic chiasm even in a single case.

Thus, although a contralateral organisation may be advantageous in some special cases, there is currently no theory to explain the extent of decussations and the extraordinary evolutionary conservation. Neither does consistent evidence exist for an evolutionary advantage of having vision, audition, somato-sensation, and motor control all contralaterally organised, in the forebrain. Furthermore, to my knowledge no existing theory explains why olfaction should be the only ipsilateral sense in the forebrain. 

The current work proposes a very different explanation, which neither leads to better visual processing nor to better motor control or sensorimotor integration. It is proposed that opposite compensatory deformations to an axial rotation in the early embryo explain the pattern of midline-crossing and non-crossing connections in the nervous system.  A model is developed to explain how these patterns can be brought about and how an optic chiasm develops as its consequence.

\section{The axial twist hypothesis} \label{secModel}

As we will see below (section \ref{secMidlineCrossings}) all extant and fossil vertebrates possess an optic chiasm. The model is therefore based on the hypothesis that an ancestor of all vertebrates has turned on its left side, by a 90\textdegree\ turn about the body axis (i.e.\ anti-clockwise from the perspective of the embryo). As the fossil record of vertebrates extends back to the cambrian, this event should have occurred at least 450 Myr ago. Here we will focus on the early embryological compensations to this side-turn and the morphological implications in extant vertebrates. Possible evolutionary scenarios will be presented in the Discussion (section \ref{secDiscEvol}). 

\begin{figure}[ht!]	 \centering
\ifthenelse{\boolean{publ}}{}{\internallinenumbers}
\includegraphics[width=7.5cm]{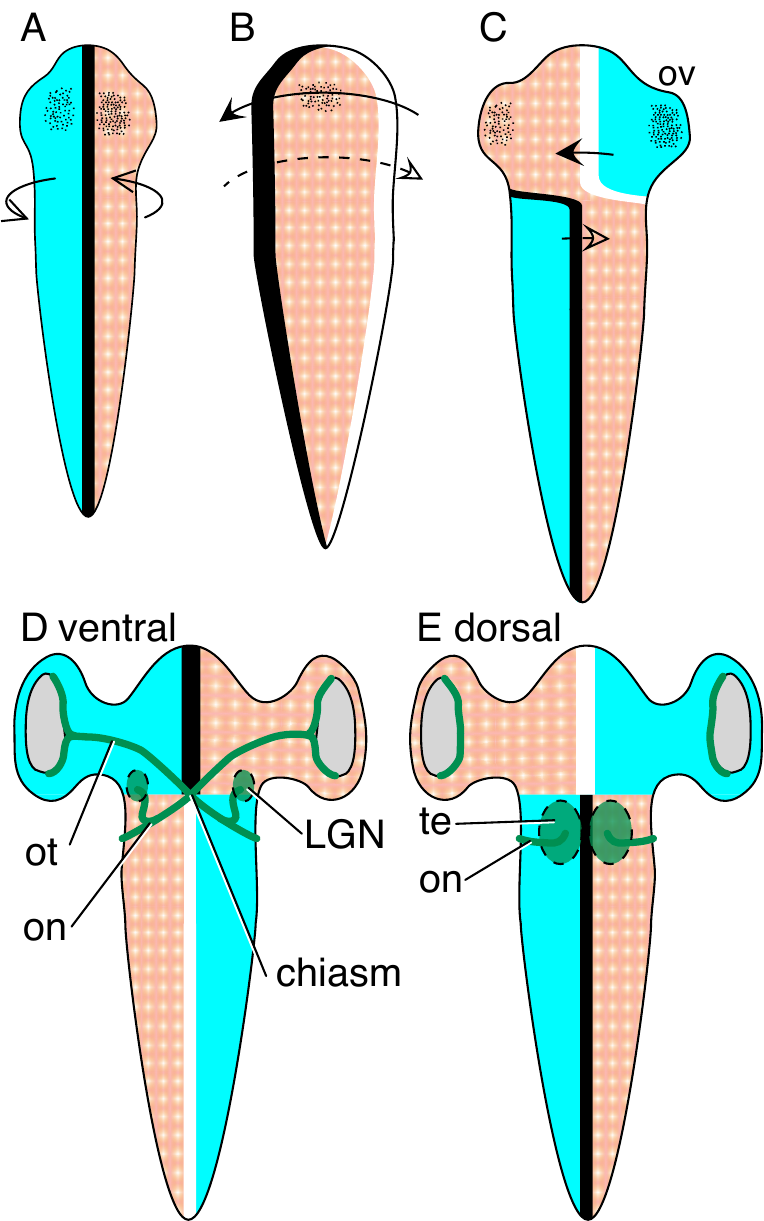}
\caption{Model of the compensations to a axial turn in the early vertebrate development; (A-C) The embryo viewed from above, with rostral up. Black zone: dorsal; white zone: ventral; spotted: right side; dotted: prospective eye region. 
The embryo turns anti-clockwise, on its left side, as indicate by the stick-arrows in (A). 
The turn is compensated by anti-clockwise (filled arrows) and clockwise (dashed arrows) movements (B). Consequently, the rostral region is inverted with respect to the rest of the body (C). The optic vesicles (ov) emerge and evaginate. 
(D-E) Development of the optic tract, in ventral view (D) and dorsal view (E). The optic tracts (ot) originate from the retinas and grow medially towards the inverted ``dorsal'' of the forebrain region (black zone in ventral view). After the chiasm the optic nerves (on) project toward the dorsal optic tectum (te) of the opposite side (panel E). Note that the optic nerves also target the lateral geniculate nucleus of the thalamus (LGN). 
} \label{fig2Model} \end{figure}

The left-turn makes that the embryo is located on its left side (fig.\ \ref{fig2Model}A-B). The body has thus lost bilateral symmetry with respect to the vertical sagittal plane. Being fishes, early vertebrates were active swimmers and so probably had a strong evolutionary advantage of a bilaterally symmetric locomotor system. By contrast, for inner body structures, such as heart and digestive organs, there is no evolutionary advantage to compensate the turn, so we expect that these will make a clear left-turn during embryology. 

There are two ways to regain bilateral symmetry. Compensations opposite to the turn (i.e.\ clockwise) will bring the body parts back into their original symmetric position. On the other hand, compensations in the same direction of the turn (i.e.\ anti-clockwise) will bring the body parts into a symmetric position that is inverted, both on the left-right and the dorso-ventral body axes, with respect to the original position. 

We propose that developmental compensations have evolved in both directions simultaneously, leading to a twist in the nervous system (fig.\ \ref{fig2Model}B-C). In figure \ref{fig2Model}B the embryo is turned on its left side. The rostral head region compensates in the direction of the turn, 90\textdegree\ anti-clockwise, so the anterior head region becomes inverted (cf.\ \ref{fig2Model}A and C). The more caudal regions, including the midbrain and the mouth, compensate against the direction of the body-turn, 90\textdegree\ clockwise  (fig.\ \ref{fig2Model}B-C). As a result, the rostral head region in the adult vertebrate is twisted and inverted with respect to the caudal body parts. 

The embryological model thus has three components. (1) A 90\textdegree\ anti-clockwise turn, (2) 90\textdegree\ anti-clockwise compensations in the rostral head region and (3) 90\textdegree\ clockwise compensations elsewhere. Note that the anti-clockwise turn may be difficult to discern from the compensatory movements, since vertebrate embryos are not free-swimming but attached to a large yolk mass.

The optic tracts develop once the retinas have been formed. The axons of the optic tract must find their way to the brain. It is known that the direction of growth of the axons is guided by a chain of molecular markers \citep[e.g.][]{lee2004.947-960,hammerschmidt2003.128-133,jeffery2005.721-753}. This guidance problem is drawn schematically in figure \ref{fig2Model}D. The eyes and forebrain are inverted, whereas the target, the optic tectum, is not. As a result, the primary molecular markers will have inverted along with the forebrain and therefore lead the tracts in a medial ventral direction. Since the tectum is not located ventral but dorsal and on the opposite side, the axons are subsequently guided across the midline toward the dorsal, contralateral tectum (fig.\ \ref{fig2Model}D).
The second target of the optic nerves is the thalamus, contralateral to the retina. The thalamic branch splits off behind the optic chiasm. 

In the following sections we will show that: 
1. 
The morphology of the forebrain is consistent with a dorsoventral inversion.
2.
The hypothesis better explains the pattern of midline crossings and ipsilateral connections in the central nervous system than does any of the existing theories. 
3. 
The axial twist model is consistent with recent observations of the early development of vertebrates: zebrafish and chick. 
4. 
The model also explains morphological peculiarities such as the asymmetric inner organs and the decussation of the trochlear nerve.
5. 
The twist is predicted by a plausible evolutionary scenario which assumes that the common ancestor of vertebrates turned on its left side and compensated for this in its further development (analogous to some flatfishes).

\section{Dorsoventral inversion of the forebrain}

The forebrain of vertebrates is in many ways a puzzling structure and is often thought to have evolved independently of the rest of the brain (for example when comparing vertebrates and cephalochordates (the lancelet)). The axial twist hypothesis provides a simple explanation if the forebrain is inverted dorsoventrally with respect to the mid- and hindbrain. 

\begin{figure}[htb!] \centering
\ifthenelse{\boolean{publ}}{}{\internallinenumbers}
\includegraphics[width=7.5 cm]{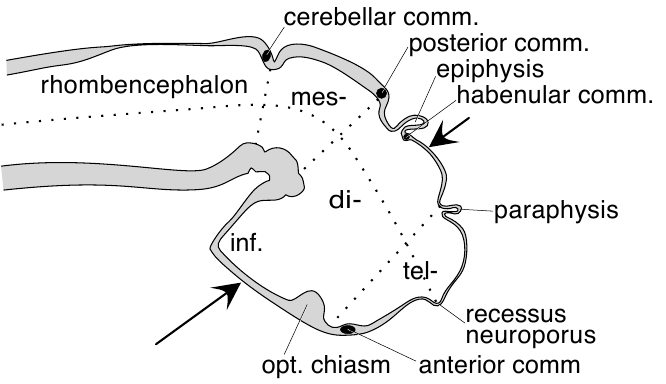}
\caption{Generalised schema of the 5-vesicle embryological stage of the vertebrate brain in medial view. The cerebellar, posterior and habenular commissures are located dorsally, whereas the anterior commissure is located ventral from the longitudinal axis of the brain (dotted line). Arrows: hypothetical location of the twist. inf: infundibulum. Redrawn from \citet{von-kupffer1906.1-272}, figure 13.}
\label{fig3commiss} \end{figure}

Figure \ref{fig3commiss} shows a generalised schema of the five-vesicle embryological stage of the vertebrate brain \citep{von-kupffer1906.1-272,nieuwenhuys1998.159-228}. There are four commissures, the cerebellar, posterior and habenular commissures are located dorsally, whereas the most rostral, the anterior commissure, is located ventrally. Each of the four commissures is associated with a sensory centre in one of the main subdivisions of the brain. The anterior commissure is associated with the olfactory tract, which is the only sensory system in the central nervous system that targets a ventral brain region. The afferent spinal nerve roots target the dorsal lobes of the spinal chord (cf.\ fig.\ \ref{figCajal}), the sensory ganglia in the medulla and pons are located dorso-laterally, the cerebellum and tecti mesencephali are located dorsally. The visual tract is a transitional structure since its rostral target, the thalamic LGN, is ventral, whereas its more caudal target, the optic tectum, is located in the dorsal midbrain.

Consequently, the gross anatomy of the forebrain is clearly consistent with an anterior dorsoventral inversion. The transitional twist is rostral from the habenular commissure and caudal from the optic chiasm (arrows in fig.\ \ref{fig3commiss}).

\section{Midline crossings and ipsilateral connections in the nervous system} \label{secMidlineCrossings}

The pattern of contralateral and ipsilateral representations in the brain is complicated and needs a more detailed treatment. First we should mention, however, that there are many bilateral connections in the central nervous system, between homologous regions on both sides (for example, the four major commissures in the brain, cf.\ figure \ref{fig3commiss}), as well as for long-range connections (for example the projections of the thalamo-spinal projections in the lamprey, see Discussion). Such bilateral connections have important functions, as mentioned in the Introduction (section \ref{secIntro}). 

In all vertebrates, extant and fossil, the \emph{tractus} of the olfactory bulb is the first cranial nerve, n.I (and probably n.0, the \emph{terminalis} as well), which connects to the telencephalon \citep{nieuwenhuys1998.book, janvier1996.book}. The olfactory tract never crosses, so smell is an ipsilateral sense. This is consistent with the hypothesis telling that both, olfaction and the forebrain, are inverted.

For the visual sense, the situation is more complex. In all vertebrates, the retina is the origin of the second cranial nerve, the optic tract (n.II). In all vertebrates the optic nerves cross or merge in the optic chiasm. The latter is located ventral to the brain, or, in cyclostomes (lamprey and hagfishes) in the ventral part of the midbrain \citep{nieuwenhuys1998.book,ronan1998.chapter}. As mentioned above, frontal-eyed vertebrates often are specialised for binocular vision. In those cases, each side of the brain receives projections from the contralateral visual hemifield of both eyes, rather than only from the contralateral eye. Otherwise, the vast majority of optic tract fibers projects to the contralateral side of the brain. 

The optic tracts project to the optic tectum (=superior colliculus in mammals) of the dorsal midbrain from the contralateral eye (or the contralateral visual hemifield). This is consistent with the hypothesis, because the eyes are inverted but the midbrain is not. In many vertebrates the  telencephalon receives visual inputs. These inputs can come from the optic tectum, or from the lateral geniculate nucleus (LGN) of the thalamus. In mammals and birds, the forebrain is predominant in visual processing. Visual fibers enter directly from the optic chiasm into the contralateral LGN \citep{nieuwenhuys2002.257-270}. Since the thalamus is part of the forebrain, the projections from the thalamus to the visual cortex do not cross the midline so that the visual cortex receives inputs from the contralateral eye (or contralateral visual hemifield). 

In sharks a large part of the telencephalon, the central nucleus, has a visual function \citep{luiten1981.539-548}. Interestingly, the central nucleus of nurse sharks represents the ipsilateral eye, as was found in electrophysiological experiments \citep{cohen1973.492-494}. This remarkable difference from other vertebrates is nevertheless consistent with the axial twist hypothesis. The central nucleus gets its input predominantly from the optic tectum, passed via thalamic nuclei \citep{smeets1981.13-23,smeets1981.1-11}. \citet{luiten1981.531-538,luiten1981.539-548} showed that nerve fibers decussate between the tectum and the thalamus. Thus, in accordance with our hypothesis, the visual information crosses the midline twice: first in the optic chiasm and again in the ventral mesencephalic tegmentum \citep{luiten1981.539-548, ebbesson1971.254-256}. 

Thus, whether the hemispheres of the telencephalon represent the ipsilateral or the contralateral eye depends on whether the visual input comes mainly from the optic tectum (chondrichthyans) or from the LGN (mammals and birds). Note, that none of the existing theories, not even Cajal's, can explain the ipsilateral representation (with two decussations) in shark brains.

For the auditory system, projections from the rhombencephalic auditory centres (cochlear nuclei, superior olivary complex, nuclei of the lateral lemniscus) to the temporal auditory cortex predominantly cross the midline, to the contralateral hemisphere \citep{loo2009.e7396,heffner1989.789-801}. This is as predicted by our hypothesis.

\section{The early development of vertebrates}

\begin{figure*}[htb!] \centering
\ifthenelse{\boolean{publ}}{}{\internallinenumbers}
\includegraphics[width=12 cm]{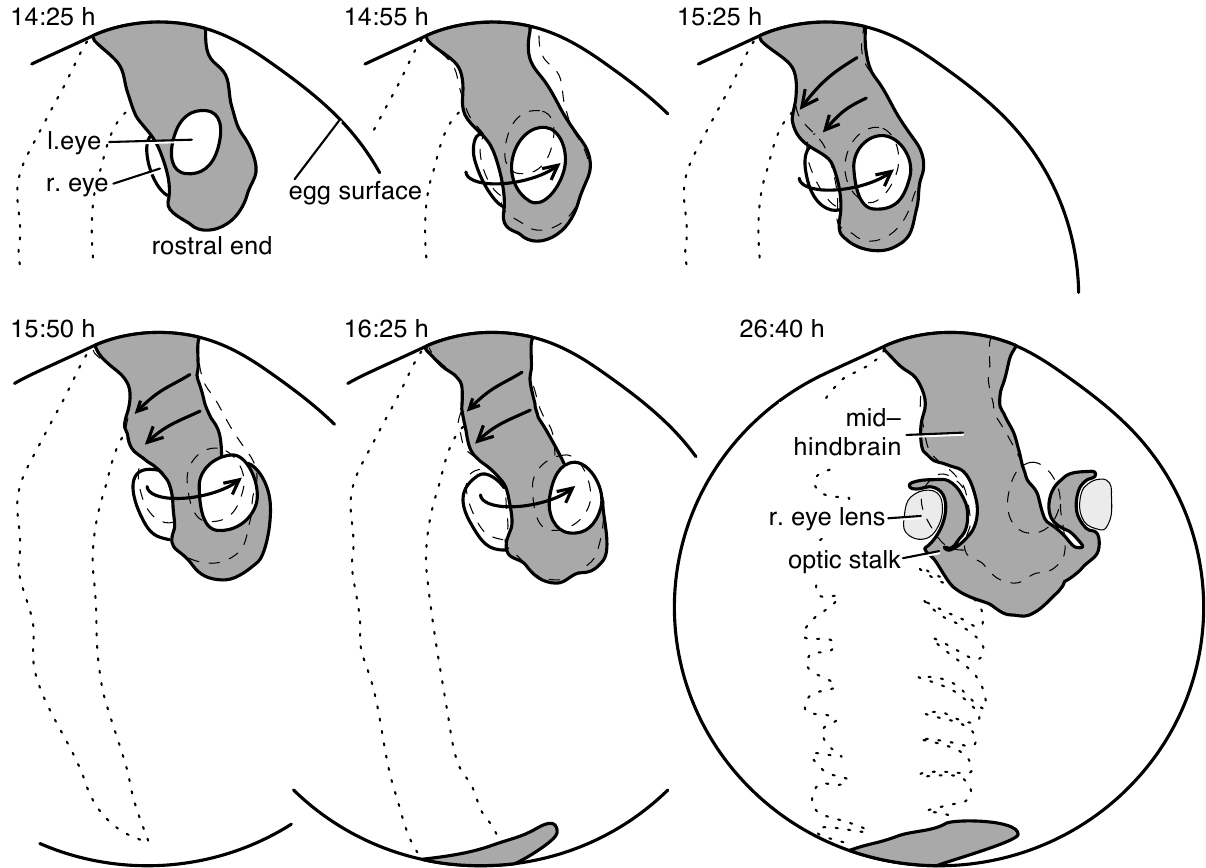}
\caption{Antero-dorsal view on the head and anterior trunk region of a zebrafish embryo (\emph{Danio rerio}) on the egg surface (see supplementary movies \ref{supp1}, \ref{supp2}). The embryo is drawn in grey, the prospective eye regions white. Dashed contours show the previous location of the embryo. The location of the body on the back side of the egg is drawn dotted. Compensatory movements can be observed between 14:40 and 16:40 p.f. During this period those cells that will form the eyes migrate anti-clockwise (perspective of the embryo), whereas the future mid- and hindbrain cells migrate clockwise between 15:15 and 16:40 h (arrows). The right eye is initially invisible because it is hidden below the cells that will form the forebrain. The first 5 frames are interleaved by 30 min, the last one is 10:15 h later. Drawn from \cite{keller2008.1065-1069}: supplementary movie no.\ 2.}
\label{fig4Danio} \end{figure*}

As explained above (section \ref{secModel}) the model distinguishes three  components: the anti-clockwise turn, the anterior anti-clockwise compensations and the more caudal clockwise compensations (fig.\ \ref{fig2Model}). Note that the model does not prescribe a temporal order for these occurrences: if the developmental processes would occur in the free-swimming stage, the larva should be adapted to swimming during each phase. The early vertebrate embryos do not swim nor move though. Being fed by a large yolk mass, the selective pressures for the temporal order in which the three processes occur presumably are defined by the level of symmetry obtained for the sensory and locomotor systems in the first free-swimming or free-moving stage.
Thus, it can be expected that the opposite clockwise and anti-clockwise compensations in the head region might be the first to develop because these involve heavy twisting after which the anterior head region is effectively inverted. 

Compensatory movements must not involve the heart and other inner organs. This is because these are not part of sensory systems or the locomotor system and therefore there is no selective pressure that these organs should be symmetric in the body. Consequently, the clockwise compensations may manifest themselves as an anti-clockwise turn of the bowels with respect the body. 

Such anti-clockwise turning movements of inner organs do indeed occur in the early embryological development of vertebrates and involve the heart, stomach, gut, liver and pancreas. This developmental turning of viscera presumably occurs in all vertebrates. Detailed descriptions exist for teleost fish (e.g.\ \citealp{barth2005.844-850}) and tetrapods (e.g.\ \citealp{hamburger1992.231-272}).

Brain development in teleosts such as the zebrafish, \emph{Danio rerio}, depends critically on the \emph{organiser} (in birds and mammals known as Hensens' node). A group of cells in the prospective anterior head region determines the development of the forebrain \citep{houart1998.788-792}. Recently, technical advances have enabled the continuous three-dimensional tracing of large numbers of cells in the live, developing teleost embryo \citep{rembold2006.1130-1134, england2006.4613-4617, keller2008.1065-1069}. Ultimately, \citet{keller2008.1065-1069} traced the first 1600 min (26:40 h) of all cells of a zebrafish. Some of the movies show the movements of a group of cells that will form the left eye (cf.\ their supplementary movie 11). Keller et al.\ did not describe the movements of the cells of the prospective right eye, and did consequently not report a turning movement of the anterior head region.

A detailed inspection of the published movies (\href{http://www.embl.de/digitalembryo/fish.html}{www.embl.de}) reveals that almost all cells that will form the right eye are invisible until 880 min
(14:40 h) after the first cell division, because they are hidden behind (towards the yolk) the prospective forebrain cells. Between 880--1000 min (14:40--16:40 h), these cells appear (fig.\ \ref{fig4Danio} and supplementary material, movies \ref{supp1} and \ref{supp2}). During this period, the cell populations that will form the two eyes move by 90\textdegree\ about the body axis. Simultaneous with the rostral movements, cells of the later mid- and hindbrain move rightward, in the opposite direction to the forebrain region (fig.\ \ref{fig4Danio}). This pattern is fully consistent with the predictions of the model for compensations of an axial turn (fig.\ \ref{fig2Model}B-C). At present, the authors have no information as to when and how the left-turn occurs in the zebrafish embryo.

\begin{figure*}[ht!] \centering
\includegraphics[width=12 cm]{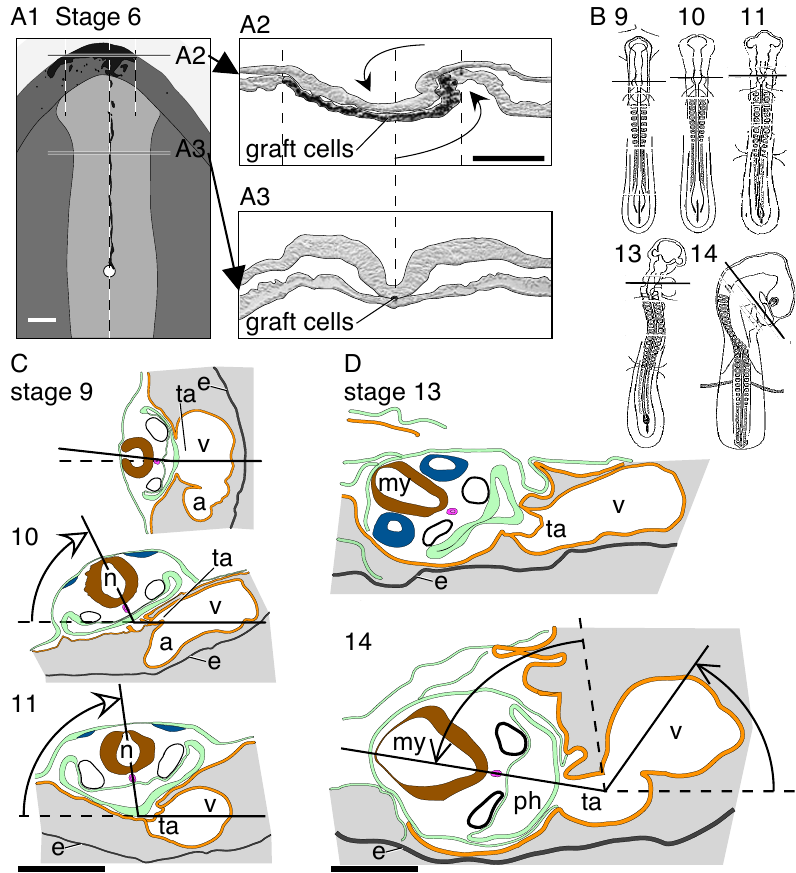}
\ifthenelse{\boolean{publ}}{}{\internallinenumbers}
\caption{
Stages of the early embryological development of the chick.
(A) Anti-clockwise (embryo's perspective) compensatory movements in the rostral region between stage 4 and 8 (presented is stage 6). A1: schematic dorsal view of the embryo. The white disk shows the location of a quail graft, implanted 24 h before (stage 3b) at the rostral end of the primitive streak. The dark spots are marked quail cells. The horizontal bars show the locations of transverse sections A2 and A3. Dashed lines mark the primitive streak (midline) and the borders of the region with quail cells in the rostral section (A2), used for alignment of the sections with the dorsal view. A2, A3: the two layers depict ectoderm and endoderm, with quail cells in black. Arrows: hypothetical anti-clockwise movements of rostral head region. 
(B) Overview of the sections in the stages 9, 10, 11, 13 and 14 of panels C and D. 
(C) Clockwise compensatory movements during stage 9-11. The ventricle (v) and \emph{truncus arteriosus} (ta) are drawn in the same orientation in each stage, whereas the embryo migrates around this orientation (arrow). The atrium (a) is located in a different section in after stage 10. Shown are the neural tube (n) flanked by the otic vesicles (from stage 10), the notochord and pharynx (ph), flanked by the two dorsal aortas. e = endoderm. 
(D) During stage 13-17 a left-turn of the body occurs (presented are stage 13 and 14). As shown by the arrows, the heart also turns. 
Panel A: adapted from \citet{lopez-sanchez2001.334-346}. Panel B-D: adapted from \citet{bellairs2005.book}. Scale bars are 300 $\mu m$.
} \label{fig5Chick} \end{figure*}

A traditionally well-studied tetrapod embryo is the chick (\emph{Gallus gallus domesticus}). The early development is usually distinguished into standardised stages \citep{hamburger1992.231-272,bellairs2005.book}. A left-turn occurs between stage 13-17, starting at the head and ending at the tail. From then on, the chick lies with its left side towards the yolk sac. Since this movement involves the entire body, including the heart, it really is a turn and not a series of compensatory movements (fig.\ \ref{fig5Chick}D). We hypothesise that this left turn is indeed the 90\textdegree\ anti-clockwise turn of the model (fig. \ref{fig2Model}A-B). Is it possible to identify the compensatory deformations as well?

It has been known for almost a century \citep{cooke2004.413-421} that a transient asymmetric development already occurs at a very early stage \citep{cooke1995.681}. The chick embryo develops perfectly symmetrically up to stage 3. At that stage, Hensen's node has formed, and a small group of cells from the rostral end of the node moves rostrally to the head process \citep{duband1982.337-350}. \citet{lopez-sanchez2001.334-346} found that cells from Hensen's node migrate anteriorly toward the rostral endoderm between stage 4 and 8, where they form a ``fountain-like'' pattern (fig.\ \ref{fig5Chick}A). This pattern is highly asymmetrical to the left of the midline. More recently, \citet{kirby2003.175-188} studied a slightly earlier stage where they found that this asymmetric distribution is not yet restricted to the rostral ``fountain'' but involves the entire forehead region. As a result, Hensen's node transiently opens to the left in stage 8 \citep{cooke1995.681, cooke2004.413-421}. These rostral asymmetric movements, which occur between stage 4 and 8, match well with the hypothetical anti-clockwise compensations (arrows in fig.\ \ref{fig5Chick}A1) as envisioned by the model (fig.\ \ref{fig2Model}B-C). These movements only involve the rostral head region and therefore are not a part of the anti-clockwise whole-body turn which starts much later, in stage 13.

Asymmetric movements have been observed in the chick embryo between stage 10 and 11 (fig.\ \ref{fig5Chick}C). In stage 9, the heart begins to develop in the body midline. During stage 10-11 the heart turns anti-clockwise as predicted. As explained above, this turn of the heart reflects the clockwise, compensatory movements indicated by the arrows in figure \ref{fig5Chick}C.

The interpretation that the turn of the heart reflects compensatory movements of the body is not in contradiction with the possibility that an asymmetric heart (and bowels) might be advantageous \citep{cooke2004.413-421}. On the contrary: had the asymmetric position in the body been disadvantageous, then the heart would have evolved to compensate along with the surrounding tissue at least in some lineages.  

We have now seen that the embryology of one teleost and one tetrapod are consistent with the predictions of the axial twist hypothesis. 
The compensatory movements of the rostral head region in the chick will have to be confirmed experimentally, for example by three-dimensional tracing techniques \citep{keller2008.1065-1069}. Also, for the zebra-fish future analyses of the individual cell movements that will give rise of the two eyes as well as the compensatory movements of the individual cells in the more caudal region should be quantified.  
Since these predictions are very specific, these observations present strong evidence for the hypothesis. Although some the observations are well-known and well-documented in the literature, there is to our knowledge no alternative theory that predicts them.

\section{The decussation of the trochlear nerve}

In addition to the optic tract there is another cranial nerve that decussates, the trochlearis nerve. Most vertebrates have six muscles to move the eye. These muscles are innervated by cranial motor nerves III (oculomotorius), IV (trochlearis) and VI (abducens). Four of the muscles are innervated by the ipsilateral oculomotorius, one by the ipsilateral abducens nerve (which originates from the hindbrain). The superior oblique muscle is the only one innervated by the contralateral trochlearis motor nerve.

This control of the eye movements is remarkable, because it is highly conserved and because it displays a complex mixture of ipsilateral and contralateral innervation of eye muscles, so that the eyes are controlled by both sides of the mid- and hindbrain \citep{fritzsch1990.491-506}. Nevertheless, eye movements are among the fastest and most accurate that vertebrates can make.

The trochlearis is the only motor nerve with a complete chiasma%
\footnote{CORRECTION MADE ON 10 MAY 2014: The original version erroneously stated ``that decussates'', which is wrong because the branch of the oculomotor nerve (III) that innervates the superior rectus eye muscle does decussate. We thank J. Voogd for kindly pointing this out \citep[see Fig.\ 17.8 of][]{nieuwenhuys2008.book}.\label{fnoteCorr}}%
, and it is the only one to leave the brainstem dorsally, even in the lamprey \citep{larsell1947.447-466} and hagfish. This conservation is remarkable because the antagonistic inferior oblique muscle is innervated by the ipsilateral oculomotorius nerve. When \citet{fritzsch1990.129-134} cut the trochlearis in \emph{Xenopus} embryos, the nerve regenerated, but with a tendency to innervate the \emph{ipsilateral eye} without any observable behavioural consequences. Also, in mice with a defective netrin receptor Unc5c, the trochlear will partly or fully innervate the \emph{ipsilateral} superior oblique muscle \citep{burgess2006.5756-5766}. Thus, there does not seem to exist a functional advantage for the decussation of the trochlear nerve. 

Nevertheless, the decussation of the trochlearis nerve never disappeared in evolution. Instead, the trochlearis nerve leaves the brain contralaterally and dorsally at the isthmus rhombomere, just as if its insertion were turned upside-down. In this property, the trochlear shows an intriguing parallel to the optic tract. Whereas the latter originates from the dorsal tectum and decussates ventrally, the trochlear originates from the ventral midbrain, to decussate dorsally. 

The explanation provided by the axial twist hypothesis is similar to that for the optic chiasm (see section \ref{secModel}). In the non-twisted vertebrate ancestor the trochlearis nerve will have innervated a muscle in the rostral head region, probably associated with the eye. After the adaptation to the left-turn the target of this trochlear motor nerve has twisted along with the eyes, and is no longer dorsal and ipsilateral, but contralateral. 

An interesting question is why the other two nerves, III and VI, do not decussate. The nucleus of the oculomotorius (n.\ III) is located slightly more rostral than that of the trochlearis (n.\ IV), but still in the midbrain and more caudal than the twist. A possible explanation is that the superior oblique muscle, that is innervated by the trochlearis nerve, is the only muscle that was already associated with the eye before the axial twist evolved. In this view the target muscles of the abducens and oculomotor nerves were originally not located in the rostral head region, but evolved as eye muscles later.


\section{Discussion}

The present paper introduced a new hypothesis to explain why vertebrates possess an optic chiasm and why the forebrain is organised contralaterally, except for olfaction. On the basis of the hypothesis, an embryological model was developed. It was shown that the dorso-ventral structure of the brain as well as the pattern of decussating and non-decussating connections between the forebrain and other regions and with the eyes is consistent with the hypothesis. On the basis of the model, predictions were made for the early development of vertebrate embryos, which were tested for the zebrafish and the chick. Finally, it was shown that even the decussation of the trochlearis motor nerve can be explained with the model. 

The discussion will treat decussations and chiasms in other brain regions and in non-vertebrate animals. The possible evolutionary scenario of the axial twist will be discussed. In the outlook we will briefly treat developmental malformations and molecular evidence.

\subsection{Other decussations and chiasms}

Vertebrates are not the only animals with decussating connections and the axial twist hypothesis does not explain all decussating connections in vertebrates. 

The optic nerve of insects has two so-called chiasms. However, these chiasms invert the order of the fibres within each optic tract rather than crossing the midline of the body. Thus each side of the insect brain processes predominantly the ipsilateral eye.

Insects like most invertebrates possess a ventral nervous system structured like a rope ladder. The central nervous system of insects does have many bilateral connections, commissures, between the bodysides. The growth mechanisms of bilateral connections have been studied in detail in \emph{Drosophila} and shows many parallels to vertebrates \citep{evans2010.79-85}. In an orderly temporal sequence of attraction and repulsion axons grow towards the midline and beyond. The decussations of such neurons differ from the decussations that are explained by the axial twist hypothesis. These axons do not grow in rostral or caudal axial directions but medially. Once they have crossed the midline they may commence to grow rostrally or caudally. This is  different from the decussations that are due to the axial twist: the axons (for example of the pyramidal tracts) at first grow in an axial direction before they cross the midline. 

Neurons in vertebrates that decussate in a manner similar to the decussating neurons of \emph{Drosophila}, are the spinal and reticular M{\"u}ller and Mauthner cells of lamprey \citep{rovainen1967.1024-1042}. The axons of these cells decussate within the segment of their origin and continue on the opposite side in rostral (M{\"u}ller) or caudal (Mauthner) directions. It has been proposed that these neurons may be homologous to the Mauthner cells of teleosts \citep{kimmel1982.112-127}. Mauthner cells in teleost fish are reticulospinal cells, developing in the reticular formation \citep{kimmel1982.112-127}. Teleost fish usually possess just a single pair of these huge cells which drive the C-start escape responses. 

Unilateral high-frequency electrical stimulation to the ventral thalamus of lampreys (\emph{Lampetra fluviatilis}) evokes alternating rhythmic activity in the left and right ventral spinal roots \citep{el-manira1997.603-616}. One might expect that the stimulation arrives slightly earlier on the ipsilateral side than contralateral to the stimulation, but this is not the case. The contralateral reaction of the spinal root leads the ipsilateral side by about 400~ms. Application of strychnine to the rhombencephalic reticular nucleus speeds up the oscillation \citep{el-manira1997.603-616}, but the contralateral side still leads by about 35~ms. Similarly, stimulation of contralateral optic tract evokes a response in the middle rhombencephalic reticular nucleus after 11 ms contralaterally and 38~ms ipsilaterally \citep[intracellular recording]{zompa1996.221-227}. These results are as predicted by the axial twist hypothesis. Due to the twisted forebrain, the contralateral thalamo-spinal and retino-rhombencephalic connections behave as if they connect the ipsilateral bodyside, whereas the ipsilateral connections are delayed to compensate the twisted morphology.

\subsection{Evolution} \label{secDiscEvol}

An important question is the evolutionary origin and survival of the axial turn that eventually evoked the twisting compensations mentioned earlier. Did the twist evolve in an early vertebrate? In a chordate, or even an early deuterostome? 

Our two most distant living vertebrate relatives are the jawless lampreys and hagfish. Adult hagfish \citep{ronan1998.chapter} and lampreys \citep{nieuwenhuys1998.397-495} both possess an optic chiasm. The chiasm in lampreys and hagfish differs from that of gnathostomes in that it is located inside the ventral brain. In \emph{Lampetra fluviatilis}, the majority of rostrally projecting spinal chord neurons have decussating axons \citep{tang1979.629-645}. Similarly, in hagfish 90\% of the optic tract neurons decussate to the contralateral optic tectum \citep{ronan1998.chapter, kishida1987.303-310, wicht1990.315-328}. The diencephalon receives bilateral visual inputs \citep{wicht1990.315-328}. Nevertheless, the general structure and connections of the forebrain of both, lampreys and hagfish, is clearly contralateral as in gnathostomes. Besides, Janvier \citep{janvier1996.book,janvier1991.567-576} presented evidence for a left-sided heart in osteostracans, a group of early, jawless vertebrates, similar to the left-sided heart of extant vertebrates (cf.\ fig.\ \ref{fig5Chick}). Thus, the conclusion that the axial twist evolved in a common ancestor of all extant vertebrates, including lampreys and hagfish, seems justified. 

\begin{figure*}[ht!]	 \centering
\ifthenelse{\boolean{publ}}{}{\internallinenumbers}
\includegraphics[width=9 cm]{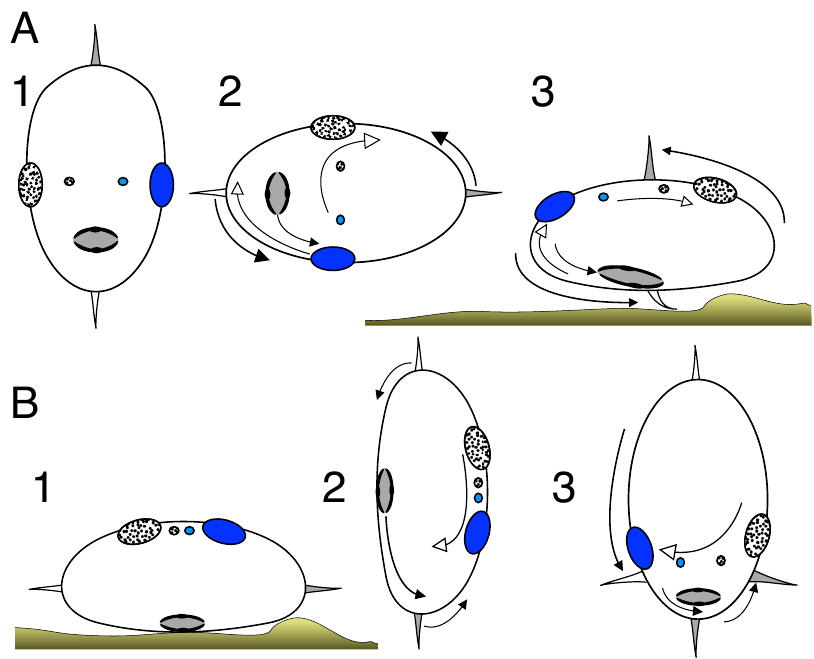}
\caption{Possible evolutionary scenarios. Panels show schematic frontal views displaying eyes, nostrils, mouth, and fin rays. One eye and one nostril are dotted to show that they do flip bodysides. 
(A) A deep bodied, free-swimming, early vertebrate (1) turned on its left side (2), for example to hide on the sea floor like a flatfish. The direction of compensation (3) of the mouth and external body parts (e.g.\ tail fin and dorsal and anal fins) is clockwise (perspective of the animal), as indicated with filled arrows. The eyes and the nostrils migrate in the opposite, anti-clockwise, direction (open arrows). 
(B) A benthic animal (1) might have turned on its left side to swim stretches or to capture prey (2). A more active form specialised on the locomotory orientation compensated the turn, again with the eyes and nostrils compensating anti-clockwise (open arrows), and the mouth and fins compensating clockwise (closed arrows, panel 2-3). 
} \label{fig6evol} 
\end{figure*}

The apparent ancientness of the axial twist places enormous difficulty on the question after the original mechanism. Why did a common ancestor of the extant vertebrates turn on its side? There are at least two conceivable scenarios. A deep-bodied animal may have turned on the side, with some resemblance to the evolution of extant flatfishes. Alternatively, a flattened animal may have turned towards an upright position (fig.\ \ref{fig6evol}).

The first scenario (fig.\ \ref{fig6evol}A) requires that an early, deep-bodied (laterally flattened) clade of vertebrates has existed which was active, free-swimming. It should have possessed paired eyes and nostrils, but no paired appendages. This animal may have become flat-bodied to hide on the sea floor, in order to overcome predation. Crucially, the deep-bodied ancestor must have had its eyes and nostrils below the body axis, because only then would these compensate in the direction of the turn and land on a crossed position, as in figure \ref{fig6evol}A3.\footnote{This can be understood if one imagines an ancestor which had its eyes above the midline. In that case, the shortest distance up for the left eye would be to migrate clockwise like the mouth and no twist would have evolved.} Thus, the first scenario predicts that the ancestor was an active animal that foraged by sight and smell from the sea floor and therefore had these senses placed ventrally or ventro-laterally, close to the mouth. It might have rested and hidden on its left side, lying on the sea floor.

The second scenario (fig.\ \ref{fig6evol}B) is complementary to the first. A flat, bottom dwelling animal might have turned on its left side. This might have been advantageous for anguiliform swimming close to the seafloor, or for capturing swimming prey or even for filter feeding. In becoming more active, it might have abandoned the bottom-dwelling orientation altogether. Again, to evolve a crossed position of the eyes and nostrils requires that the resulting, deep-bodied form had eyes and nostrils below the body axis. This scenario would be consistent with a theory by Dzik, who found that an early cambrian animal, \emph{Yunnanozoon lividum}, was possibly a vertebrate with bilateral ventro-rostral eyes \citep{chen1995.720-722}. According to \citet{dzik1995.341-360}, the precambrian ancestor of Yunnanozoon might have been a precambrian Dickinsonia. The ventro-rostral position of the presumed eyes and the deep-bodied form of \emph{Yunnanozoon}, and the flattened bottom-dwelling dickinsonians would match well with the axial twist hypothesis. 

The axial twist hypothesis is not consistent with the calcichordate hypothesis \citep{jefferies1986.book}, which proposes that a sessile ancestor (\emph{Dexiothetica}) of solutes \citep[an extinct clade of blastozoan echinoderms][]{david2000.529-555} lied down on its right side and gradually became more symmetric via cephalochordates to vertebrates. Jefferies' phylogeny \citep{jefferies1997.1-10} is not supported by phylogenetic and fossil evidence \citep{lefebvre2005.477-486, david2000.529-555}. Although echinoderm embryos do turn to the left, their complex twisting is very different to that in vertebrates \citep{lussanet2011.echino}.

The asymmetrically developing Cephalochordates (such as \emph{Amphioxus}) are more closely related to vertebrates than are echinoderms \citep{bourlat2006.85-88,delsuc2006.965-968}. The mouth of cephalochordates develops at first on the left side of the body before it migrates (anti-clockwise) to the ventral position of the adult. Note that this is the inverse of vertebrates. Cephalochordates only possess median, non-paired, visual and olfactory organs (small cell clusters) and show no signs of a twist \citep{lacalli2004.148-162}. However, since cephalochordates are laterally flattened, the mouth appearing on the left side and the first gill slits on the other, they match well with Dzik's Dickinsonia hypothesis, and so with the scenario of figure \ref{fig6evol}B. 

Thus, although at present the evolutionary mechanism behind the axial twist hypothesis is open, there are at least two plausible mechanisms in accordance with the observations from different sources including morphology, embryology and paleontology. Of these two, one bears resemblance to the evolution of extant flatfishes, whereas the latter is backed by an interesting proposal of \citet{dzik1995.341-360} and might provide a link to the asymmetric development of lancelets, as explained above.

\subsection{Outlook}

The evolution and development of body asymmetries and left-right and dorso-ventral patterning in the body has long intrigued biologists. This interest has strongly revived since the development of molecular techniques. Since the early works \citep{levin1998.67-76,brown1990.1-9} a large number of transcription factors have been linked with early dorso-ventral and left-right patterning, and it has become clear that these processes are guided by complex hierarchical interactions which begin very early in development (cf.\ fig.\ \ref{fig5Chick}A) and in which the organiser plays a central role \citep{hamada2002.103-113, rodriguez-gallardo2005.191-201, aldworth2011.e1002041, levin2004.197-206, lee2008.3464-3476, raya2004.1043-1054, raya2006.283-293}.

At present, no study has viewed these results from a perspective that the pattern in the anterior region is opposite to more caudal regions, as suggested by the present work. However there are results to suggest that the molecular patterning may agrees with the axial twist hypothesis. 
The homeobox transcription factor \emph{Gsh2} (which regulates \emph{Pax5} and \emph{Dbx1}) is expressed in the vertebrate dorsal neural tube and the ventral forebrain \citep{wilson2005.1-13}. \citet{hibino2006.587-595} mention that \emph{Pitx} genes play a central role in left-right patterning. \emph{Pitx2} is located on the left bodyside in birds and mammals, but on the right rostral from Hensen's node \citep{schlueter2007.256-267}. These results suggest that the axial twist model may present a useful model for analysing and interpreting the developmental patterning in vertebrates. 

The axial twist hypothesis also presents a model that explains why some developmental anomalies, such as a \emph{situs inversus viscerum} originate in the asymmetric formation of the organiser (cf.\ fig.\ \ref{fig5Chick}A) long before development of the heart. ALso, it explains why chicks that turn to the right always develop situs inversus \citep{von-baer1828.book}. 

Holoprosencephaly is a common and dramatic developmental malformation of the forebrain \citep{matsunaga1977.261-272,fernandes2008.413-423}. In severe forms, the forebrain remains severely underdeveloped. In so-called interhemispheric holoprosencephaly the cerebral hemispheres have a left-right orientation instead of the normal antero-dorsal orientation \citep{simon2002.151-155}. This orientation is consistent with a failure of the anti-clockwise compensations in the forebrain region. These examples show that the axial twist hypothesis may provide helpful in shedding new light on the mechanisms underlying common developmental malformations. 

Asymmetric development and twisting movements are common in the animal kingdom. Among the deu\-tero\-stomes, vertebrates (this manuscript), echinoderms \citep{lussanet2011.echino}, cephalochordates (see above), but also tunicates develop strongly asymmetric, whereas hemichordates seem to be the only major clade to develop symmetrically. Among other bilaterians, many molluscs twist strongly and obtain so a looped gut. In this light, the axial twist hypothesis for vertebrates is not extraordinary.

\section*{Acknowledgements}
We thank Kim Bostr{\"o}m, Jaap de Lussanet, Luciano Fadiga, Frietson Galis, Obbo Haze\-win\-kel, Philippe Janvier, Wim van de Grind, Mees Muller, and H-P Schultze, as well as a number of anonymous reviewers for their comments and discussions. One  author (MdL) is funded by the is supported by the German Federal Ministry of Education and Research (BMBF), project 01EC1003A.

\section*{References}

\begin{thebibliography}{84}
\expandafter\ifx\csname natexlab\endcsname\relax\def\natexlab#1{#1}\fi
\expandafter\ifx\csname url\endcsname\relax
  \def\url#1{\texttt{#1}}\fi
\expandafter\ifx\csname urlprefix\endcsname\relax\def\urlprefix{URL }\fi

\bibitem[{Aldworth et~al.(2011)Aldworth, Dimitrov, Cummins, Gedeon, and
  Miller}]{aldworth2011.e1002041}
Aldworth, Z.~N., Dimitrov, A.~G., Cummins, G.~I., Gedeon, T., Miller, J.~P.,
  2011. Temporal encoding in a nervous system. PLoS Comput. Biol. 7~(5),
  e1002041.

\bibitem[{Barth et~al.(2005)Barth, Miklosi, Watkins, Bianco, Wilson, and
  Andrew}]{barth2005.844-850}
Barth, K.~A., Miklosi, A., Watkins, J., Bianco, I.~H., Wilson, S.~W., Andrew,
  R.~J., 2005. {\em fsi} {Z}ebrafish show concordant reversal of laterality of
  viscera, neuroanatomy, and a subset of behavioral responses. Curr. Biol.
  15~(9), 844--850.

\bibitem[{Bellairs and Osmond(2005)}]{bellairs2005.book}
Bellairs, R., Osmond, M., 2005. The atlas of chick development, 2nd Edition.
  Academic Press, New York.

\bibitem[{Bertin(1994)}]{bertin1994.book}
Bertin, R. J.~V., 1994. Natural smartness in hypothetical animals: Of paddlers
  and glowballs. Ph.D. thesis, Universiteit Utrecht.

\bibitem[{Bourlat et~al.(2006)Bourlat, Juliusdottir, Lowe, Freeman, Aronowicz,
  Kirschner, Lander, Thorndyke, Nakano, Kohn, Heyland, Moroz, Copley, and
  Telford}]{bourlat2006.85-88}
Bourlat, S.~J., Juliusdottir, T., Lowe, C.~J., Freeman, R., Aronowicz, J.,
  Kirschner, M., Lander, E.~S., Thorndyke, M., Nakano, H., Kohn, A.~B.,
  Heyland, A., Moroz, L.~L., Copley, R.~R., Telford, M.~J., 2006. Deuterostome
  phylogeny reveals monophyletic chordates and the new phylum {X}enoturbellida.
  Nature 444~(7115), 85--88.

\bibitem[{Braitenberg(1984)}]{braitenberg1984.book}
Braitenberg, V., 1984. Vehicles-experiments in synthetic psychology. MIT Press,
  Cambridge, MA.

\bibitem[{Brown and Wolpert(1990)}]{brown1990.1-9}
Brown, N., Wolpert, L., 1990. The development of handedness in left/right
  asymmetry. Development 109~(1), 1--9.

\bibitem[{Budelmann(1995)}]{budelmann1995.115-138}
Budelmann, B., 1995. The cephalopod nervous system: what evolution has made of
  the molluscan design. In: Braidbach, O., Kutsch, W. (Eds.), The nervous
  system of vertebrates: an evolutionary and comparative approach. Birkhauser
  Verlag, Basel, pp. 115--138.

\bibitem[{Burgess et~al.(2006)Burgess, Jucius, and
  Ackerman}]{burgess2006.5756-5766}
Burgess, R.~W., Jucius, T.~J., Ackerman, S.~L., 2006. Motor axon guidance of
  the mammalian trochlear and phrenic nerves: Dependence on the netrin receptor
  {{\em Unc5c}} and modifier loci. J. Neurosci. 26~(21), 5756--5766.

\bibitem[{Chen et~al.(1995)Chen, Dzik, Edgecombe, Ramskold, and
  Zhou}]{chen1995.720-722}
Chen, J.~Y., Dzik, J., Edgecombe, G.~D., Ramskold, L., Zhou, G.~Q., 1995. A
  possible {E}arly {C}ambrian chordate. Nature 377~(6551), 720--722.

\bibitem[{Cohen et~al.(1973)Cohen, Duff, and Ebbesson}]{cohen1973.492-494}
Cohen, D.~H., Duff, T.~A., Ebbesson, S. O.~E., 1973. Electrophysiological
  identification of a visual area in shark telencephalon. Science 182~(4111),
  492--494.

\bibitem[{Cooke(1995)}]{cooke1995.681}
Cooke, J., 1995. Vertebrate embryo handedness. Nature 374, 681.

\bibitem[{Cooke(2004)}]{cooke2004.413-421}
Cooke, J., 2004. The evolutionary origins and significance of vertebrate
  left--right organisation. BioEssays 26~(4), 413--421.

\bibitem[{David et~al.(2000)David, Lefebvre, Mooi, and
  Parsley}]{david2000.529-555}
David, B., Lefebvre, B., Mooi, R., Parsley, R., 2000. Are homalozoans
  echinoderms? {A}n answer from the extraxial-axial theory. Paleobiol. 26~(4),
  529--555.

\bibitem[{de~Lussanet(2011)}]{lussanet2011.echino}
de~Lussanet, M. H.~E., 2011. A hexamer origin of the echinoderms' five rays.
  Evol. Dev. 13~(2), 228--238.

\bibitem[{Delsuc et~al.(2006)Delsuc, Brinkmann, Chourrout, and
  Philippe}]{delsuc2006.965-968}
Delsuc, F., Brinkmann, H., Chourrout, D., Philippe, H., 2006. Tunicates and not
  cephalochordates are the closest living relatives of vertebrates. Nature
  439~(7079), 965--968.

\bibitem[{Duband and Thiery(1982)}]{duband1982.337-350}
Duband, J.~L., Thiery, J.~P., 1982. Appearance and distribution of fibronectin
  during chick embryo gastrulation and neurulation. Dev. Biol. 94~(2),
  337--350.

\bibitem[{Dzik(1995)}]{dzik1995.341-360}
Dzik, J., 1995. Yunnanozoon and the ancestry of chordates. Acta Palaeontologica
  Polonica 40~(4), 341--360.

\bibitem[{Ebb\-esson and Ito(1980)}]{ebbesson1980.483-495}
Ebb\-esson, S. O.~E., Ito, H., 1980. Bilateral retinal projections in the black
  piranah ({{\em Serrasalmus niger}}). Cell Tissue Res. 213~(3), 483--495.

\bibitem[{Ebbesson(1980)}]{ebbesson1980.179-212}
Ebbesson, S. O.~E., 1980. The parcellation theory and its relation to
  interspecific variability in brain organization, evolutionary and ontogenetic
  development, and neuronal plasticity. Cell Tissue Res. 213~(2), 179--212.

\bibitem[{Ebbesson and Schroeder(1971)}]{ebbesson1971.254-256}
Ebbesson, S. O.~E., Schroeder, D.~M., 1971. Connections of the nurse shark's
  telencephalon. Science 173~(3993), 254--256.

\bibitem[{El~Manira et~al.(1997)El~Manira, Pombal, and
  Grillner}]{el-manira1997.603-616}
El~Manira, A., Pombal, M., Grillner, S., 1997. Diencephalic projection to
  reticulospinal neurons involved in the initiation of locomotion in adult
  lampreys {{\em Lampetra fluviatilis}}. J. Comp. Neurol. 389~(4), 603--616.

\bibitem[{England et~al.(2006)England, Blanchard, Mahadevan, and
  Adams}]{england2006.4613-4617}
England, S.~J., Blanchard, G.~B., Mahadevan, L., Adams, R.~J., 2006. A dynamic
  fate map of the forebrain shows how vertebrate eyes form and explains two
  causes of cyclopia. Development 133~(23), 4613--4617.

\bibitem[{Evans and Bashaw(2010)}]{evans2010.79-85}
Evans, T.~A., Bashaw, G.~J., 2010. Axon guidance at the midline: of mice and
  flies. Curr. Opin. Neurobiol. 20~(1), 79--85.

\bibitem[{Fernandes and H{\'e}bert(2008)}]{fernandes2008.413-423}
Fernandes, M., H{\'e}bert, J.~M., 2008. The ups and downs of holoprosencephaly:
  dorsal versus ventral patterning forces. Clin. Genet. 73~(5), 413--423.

\bibitem[{Flechsig(1899)}]{flechsig1899.III-VI}
Flechsig, P., 1899. Foreword [in german]. In: {Die Struktur des Chiasma optikum
  (german translation), by Ram{\'o}n y Cajal}. J.A. Barth, Leipzig, pp.
  III--VI.

\bibitem[{Fritzsch and Sonntag(1990)}]{fritzsch1990.129-134}
Fritzsch, B., Sonntag, R., 1990. Oculomotor ({N III}) motoneurons can innervate
  the superior oblique muscle of {{\em Xenopus}} after larval trochlear ({N
  IV}) nerve surgery. Neurosci. Lett. 114~(2), 129--134.

\bibitem[{Fritzsch et~al.(1990)Fritzsch, Sonntag, Dubuc, Ohta, and
  Grillner}]{fritzsch1990.491-506}
Fritzsch, B., Sonntag, R., Dubuc, R., Ohta, Y., Grillner, S., 1990.
  Organization of the six motor nuclei innervating the ocular muscles in
  lamprey. J. Comp. Neurol. 294~(4), 491--506.

\bibitem[{Hamada et~al.(2002)Hamada, Meno, Watanabe, and
  Saijoh}]{hamada2002.103-113}
Hamada, H., Meno, C., Watanabe, D., Saijoh, Y., 2002. Establishment of
  vertebrate left-right asymmetry. Nat. Rev. Genet. 3~(2), 103--113.

\bibitem[{Hamburger and Hamilton(1951)}]{hamburger1992.231-272}
Hamburger, V., Hamilton, H.~L., 1951. A series of normal stages in the
  development of the chick embryo. {[Reprinted in {\em Dev. Dyn.}}
  195(4):231--72 (1992)]. J. Morphol. 88, 49--92.

\bibitem[{Hammerschmidt et~al.(2003)Hammerschmidt, Kramer, Nowak, Herzog, and
  Wittbrodt}]{hammerschmidt2003.128-133}
Hammerschmidt, M., Kramer, C., Nowak, M., Herzog, W., Wittbrodt, J., 2003. Loss
  of maternal {{\em Smad5}} in zebrafish embryos affects patterning and
  morphogenesis of optic primordia. Dev. Dyn. 227~(1), 128--133.

\bibitem[{Hausen(1984)}]{hausen1984.523-555}
Hausen, K., 1984. The lobula-complex of the fly: structure, function and
  significance in visual behavior. In: Ali, M.~A. (Ed.), Photoreception and
  vision in invertebrates. Plenum Press, New York, pp. 523--555.

\bibitem[{Heffner and Heffner(1989)}]{heffner1989.789-801}
Heffner, H.~E., Heffner, R.~S., 1989. Unilateral auditory cortex ablation in
  macaques results in a contralateral hearing loss. J. Neurophysiol. 62~(3),
  789--801.

\bibitem[{Hibino et~al.(2006)Hibino, Nishino, and Amemiya}]{hibino2006.587-595}
Hibino, T., Nishino, A., Amemiya, S., 2006. Phylogenetic correspondence of the
  body axes in bilaterians is revealed by the right-sided expression of pitx
  genes in echinoderm larvae. Develop. Growth Differ. 48, 587--595.

\bibitem[{Houart et~al.(1998)Houart, Westerfield, and
  Wilson}]{houart1998.788-792}
Houart, C., Westerfield, M., Wilson, S.~W., 1998. A small population of
  anterior cells patterns the forebrain during zebrafish gastrulation. Nature
  391~(6669), 788--792.

\bibitem[{Janvier(1996)}]{janvier1996.book}
Janvier, P., 1996. Early vertebrates. Oxford University Press.

\bibitem[{Janvier et~al.(1991)Janvier, Percy, and Potter}]{janvier1991.567-576}
Janvier, P., Percy, L.~R., Potter, I.~C., 1991. The arrangement of the heart
  chambers and associated blood vessels in the {D}evonian osteostracan {{\em
  Norselaspis glacialis}}. {A} reinterpretation based on recent studies of the
  circulatory system in lampreys. J. Zool. 223~(4), 567--576.

\bibitem[{Jefferies(1986)}]{jefferies1986.book}
Jefferies, R. P.~S., 1986. The ancestry of the vertebrates. British Museum
  (Natural History), London.

\bibitem[{Jefferies(1997)}]{jefferies1997.1-10}
Jefferies, R. P.~S., 1997. A defence of the calcichordates. Lethaia 30~(1),
  1--10.

\bibitem[{Jeffery and Erskine(2005)}]{jeffery2005.721-753}
Jeffery, G., Erskine, L., 2005. Variations in the architecture and development
  of the vertebrate optic chiasm. Prog. Retinal Eye Res. 24~(6), 721--753.

\bibitem[{Keller et~al.(2008)Keller, Schmidt, Wittbrodt, and
  Stelzer}]{keller2008.1065-1069}
Keller, P.~J., Schmidt, A.~D., Wittbrodt, J., Stelzer, E.~H., 2008.
  Reconstruction of zebrafish early embryonic development by scanned light
  sheet microscopy. Science 322~(5904), 1065--1069.

\bibitem[{Kimmel et~al.(1982)Kimmel, Powell, and Metcalfe}]{kimmel1982.112-127}
Kimmel, C.~B., Powell, S.~L., Metcalfe, W.~K., 1982. Brain neurons which
  project to the spinal cord in young larvae of the zebrafish. J. Comp. Neurol.
  205~(2), 112--127.

\bibitem[{Kirby et~al.(2003)Kirby, Lawson, Stadt, Kumiski, Wallis, McCraney,
  Waldo, Li, and Schoenwolf}]{kirby2003.175-188}
Kirby, M.~L., Lawson, A., Stadt, H.~A., Kumiski, D.~H., Wallis, K.~T.,
  McCraney, E., Waldo, K.~L., Li, Y.-X., Schoenwolf, G.~C., 2003. Hensen's node
  gives rise to the ventral midline of the foregut: implications for organizing
  head and heart development. Dev. Biol. 253~(2), 175--188.

\bibitem[{Kishida et~al.(1987)Kishida, Goris, Nishizawa, Koyama, Kadota, and
  Amemiya}]{kishida1987.303-310}
Kishida, R., Goris, R.~C., Nishizawa, H., Koyama, H., Kadota, T., Amemiya, F.,
  1987. Primary neurons of the lateral line nerves and their central
  projections in hagfishes. J. Comp. Neurol. 264, 303--310.

\bibitem[{Lacalli(2004)}]{lacalli2004.148-162}
Lacalli, T.~C., 2004. Sensory systems in amphioxus: A window on the ancestral
  chordate condition. Brain Behav. Evol. 64~(3), 148--162.

\bibitem[{Larsell(1947)}]{larsell1947.447-466}
Larsell, O., 1947. The nucleus of the {IVth nerve in Petromyzonts}. J. Comp.
  Neurol. 86~(3), 447--466.

\bibitem[{Lee and Anderson(2008)}]{lee2008.3464-3476}
Lee, J.~D., Anderson, K.~V., 2008. Morphogenesis of the node and notochord: The
  cellular basis for the establishment and maintenance of left--right asymmetry
  in the mouse. Dev. Dyn. 237~(12), 3464--3476.

\bibitem[{Lee et~al.(2004)Lee, von~der Hardt, Rusch, Stringer, Stickney,
  Talbot, Geisler, N{\"u}sslein-Volhard, Selleck, Chien, and
  Roehl}]{lee2004.947-960}
Lee, J.-S., von~der Hardt, S., Rusch, M.~A., Stringer, S.~E., Stickney, H.~L.,
  Talbot, W.~S., Geisler, R., N{\"u}sslein-Volhard, C., Selleck, S.~B., Chien,
  C.-B., Roehl, H., 2004. Axon sorting in the optic tract requires {HSPG}
  synthesis by ext2 (dackel) and extl3 (boxer). Neuron 44, 947--960.

\bibitem[{Lefebvre(2005)}]{lefebvre2005.477-486}
Lefebvre, B., 2005. Stylophoran supertrees revisited. J. Paleontol. Polonica
  50~(3), 477--486.

\bibitem[{Levin(1998)}]{levin1998.67-76}
Levin, M., 1998. Left-right asymmetry and the chick embryo. Sem. Cell Developm.
  Biol. 9~(1), 67--76.

\bibitem[{Levin(2004)}]{levin2004.197-206}
Levin, M., 2004. The embryonic origins of left-right asymmetry. Crit. Rev. Oral
  Biol. Med. 15~(4), 197--206.

\bibitem[{Llin{\'a}s(2003)}]{llinas2003.77-80}
Llin{\'a}s, R.~R., 2003. The contribution of {Santiago Ram{\'o}n y Cajal} to
  functional neuroscience. Nat. Rev. Neurosci. 4, 77--80.

\bibitem[{Loeb(1918)}]{loeb1918.book}
Loeb, J., 1918. Forced movements, tropisms and animal conduct. Lippincott,
  republished 1973, Dover, New York.

\bibitem[{Loosemore(2009)}]{loosemore2009.375-382}
Loosemore, R.~G., 2009. The inversion hypothesis: A novel explanation for the
  contralaterality of the human brain. Bioscience Hypotheses 2~(66), 375--382.

\bibitem[{Lopez-Sanchez et~al.(2001)Lopez-Sanchez, Garcia-Martinez, and
  Schoenwolf}]{lopez-sanchez2001.334-346}
Lopez-Sanchez, C., Garcia-Martinez, V., Schoenwolf, G.~C., 2001. Localization
  of cells of the prospective neural plate, heart and somites within the
  primitive streak and epiblast of avian embryos at intermediate
  primitive-streak stages. Cells Tissues Organs 169~(4), 334--346.

\bibitem[{Luiten(1981{\natexlab{a}})}]{luiten1981.531-538}
Luiten, P. G.~M., 1981{\natexlab{a}}. Two visual pathways to the telencephalon
  in the nurse shark ({{\em Ginglymostoma cirratum}}). {I}. retinal
  projections. J. Comp. Neurol. 196~(4), 531--538.

\bibitem[{Luiten(1981{\natexlab{b}})}]{luiten1981.539-548}
Luiten, P. G.~M., 1981{\natexlab{b}}. Two visual pathways to the telencephalon
  in the nurse shark ({{\em Ginglymostoma cirratum}}). {II}. ascending
  thalamo-telencephalic connections. J. Comp. Neurol. 196~(4), 539--548.

\bibitem[{Matsunaga and Shiota(1977)}]{matsunaga1977.261-272}
Matsunaga, E., Shiota, K., 1977. Holoprosencephaly in human embryos:
  epidemiologic studies of 150 cases. Teratology 16~(3), 261--272.

\bibitem[{Nieuwenhuys(1998)}]{nieuwenhuys1998.159-228}
Nieuwenhuys, R., 1998. Morphogenesis and general structure. In:
  \cite{nieuwenhuys1998.book}, Ch.~4, pp. 159--228.

\bibitem[{Nieuwenhuys(2002)}]{nieuwenhuys2002.257-270}
Nieuwenhuys, R., 2002. Deuterostome brains: Synopsis and commentary. Brain Res.
  Bull. 57~(3--4), 257--270.

\bibitem[{Nieuwenhuys et~al.(1998)Nieuwenhuys, Donkelaar, Nicholson, Smeets,
  and Wicht}]{nieuwenhuys1998.book}
Nieuwenhuys, R., Donkelaar, H.~J., Nicholson, C., Smeets, W. J. A.~J., Wicht,
  H., 1998. The central nervous system of vertebrates. Springer, New York.

\bibitem[{Nieuwenhuys and Nicholson(1998)}]{nieuwenhuys1998.397-495}
Nieuwenhuys, R., Nicholson, C., 1998. Lampreys, {Petromyzontoidea}. In:
  \cite{nieuwenhuys1998.book}, Ch.~10, pp. 397--495.

\bibitem[{Nieuwenhuys et~al.(2008)Nieuwenhuys, Voogd,
  and van Huijzen}]{nieuwenhuys2008.book}
Nieuwenhuys, R., Voogd, J., and van Huijzen, C., 2008. The human central nervous system. Springer, Berlin, 4th edition.

\bibitem[{Ram{\'o}n~y Cajal(1898)}]{ramon-y-cajal1898.15-65}
Ram{\'o}n~y Cajal, S., 1898. Estructura del kiasma optico y teoria general de
  los entrecruzamientos de las vias nerviosas [german 1899, english 2004]. Rev.
  Trim. Microgr{\'a}fica 3, 15--65.

\bibitem[{Raya and Belmonte(2006)}]{raya2006.283-293}
Raya, A., Belmonte, J. C.~I., 2006. Left-right asymmetry in the vertebrate
  embryo: From early information to higher-level integration. Nat. Rev. Genet.
  7~(4), 283--293.

\bibitem[{Raya and Izpisua~Belmonte(2004)}]{raya2004.1043-1054}
Raya, A., Izpisua~Belmonte, J.~C., 2004. Unveiling the establishment of
  left-right asymmetry in the chick embryo. Mech. Dev. 121~(9), 1043--1054.

\bibitem[{Regan(2000)}]{regan2000.book}
Regan, D., 2000. Human perception of objects: early visual processing of
  spatial form defined by luminance, color, texture, motion, and binocular
  disparity. Sinauer Assoc., Sunderland, MA.

\bibitem[{Rembold et~al.(2006)Rembold, Loosli, Adams, and
  Wittbrodt}]{rembold2006.1130-1134}
Rembold, M., Loosli, F., Adams, R.~J., Wittbrodt, J., 2006. Individual cell
  migration serves as the driving force for optic vesicle evagination. Science
  313~(5790), 1130--1134.

\bibitem[{Rodr{\'\i}guez-Gallardo et~al.(2005)Rodr{\'\i}guez-Gallardo,
  S{\'a}nchez-Arrones, Fern{\'a}ndez-Garre, and
  Puelles}]{rodriguez-gallardo2005.191-201}
Rodr{\'\i}guez-Gallardo, L., S{\'a}nchez-Arrones, L., Fern{\'a}ndez-Garre, P.,
  Puelles, L., 2005. Agreement and disagreement among fate maps of the chick
  neural plate. Brain Res. Rev. 49~(2), 191--201.

\bibitem[{Ronan and Northcutt(1998)}]{ronan1998.chapter}
Ronan, M., Northcutt, R.~G., 1998. The central nervous system of hagfishes. In:
  J{\o}rgensen, J.~M., Lomholt, J.~P., Weber, R.~E., Malte, H. (Eds.), The
  biology of hagfishes. Chapman \& Hall, London, pp. 452--477.

\bibitem[{Rovainen(1967)}]{rovainen1967.1024-1042}
Rovainen, C.~M., 1967. Physiological and anatomical studies on large neurons of
  central nervous system of the sea lamprey ({{\em Petromyzon marinus}). II.
  D}orsal cells and giant interneurons. J. Neurophysiol. 30~(5), 1024--1042.

\bibitem[{Schlueter and Brand(2007)}]{schlueter2007.256-267}
Schlueter, J., Brand, T., 2007. Left-right axis development: examples of
  similar and divergent strategies to generate asymmetric morphogenesis in
  chick and mouse embryos. Cytogenet. Gen. Res. 117~(1-4), 256--267.

\bibitem[{Shinbrot and Young(2008)}]{shinbrot2008.1278-1292}
Shinbrot, T., Young, W., 2008. Why decussate? topological constraints on {3D}
  wiring. Anat. Record 291~(10), 1278--1292.

\bibitem[{Simon et~al.(2002)Simon, Hevner, Pinter, Clegg, Delgado, Kinsman,
  Hahn, and Barkovich}]{simon2002.151-155}
Simon, E.~M., Hevner, R.~F., Pinter, J.~D., Clegg, N.~J., Delgado, M., Kinsman,
  S.~L., Hahn, J.~S., Barkovich, A.~J., 2002. The middle interhemispheric
  variant of holoprosencephaly. Am. J. Neuroradiol. 23, 151--155.

\bibitem[{Smeets(1981{\natexlab{a}})}]{smeets1981.13-23}
Smeets, W. J. A.~J., 1981{\natexlab{a}}. Efferent tectal pathways in two
  chondrichthyans, the shark {{\em Scyliorhinus canicula}} and the ray {{\em
  Raja clavata}}. J. Comp. Neurol. 195~(1), 13--23.

\bibitem[{Smeets(1981{\natexlab{b}})}]{smeets1981.1-11}
Smeets, W. J. A.~J., 1981{\natexlab{b}}. Retinofugal pathways in two
  chondrichthyans, the shark {{\em Scyliorhinus canicula}} and the ray {{\em
  Raja clavata}}. J. Comp. Neurol. 195~(1), 1--11.

\bibitem[{Stern(2002)}]{stern2002.447-451}
Stern, C.~D., 2002. Induction and initial patterning of the nervous system --
  the chick embryo enters the scene. Curr. Opin. Genet. Develop. 12, 447--451.

\bibitem[{Tang and Selzer(1979)}]{tang1979.629-645}
Tang, D., Selzer, M.~E., 1979. Projections of lamprey spinal neurons determined
  by the retrograde axonal transport of horseradish peroxidase. J. Comp.
  Neurol. 188~(4), 629--645.

\bibitem[{van~der Loo et~al.(2009)van~der Loo, Gais, Congedo, Vanneste,
  Plazier, Menovsky, Van~de Heyning, and De~Ridder}]{loo2009.e7396}
van~der Loo, E., Gais, S., Congedo, M., Vanneste, S., Plazier, M., Menovsky,
  T., Van~de Heyning, P., De~Ridder, D., 2009. Tinnitus intensity dependent
  gamma oscillations of the contralateral auditory cortex. PLoS ONE 4~(10),
  e7396.

\bibitem[{von Baer(1828)}]{von-baer1828.book}
von Baer, K.~E., 1828. {\"U}ber {E}ntwickelungsgeschichte der {T}hiere.
  {B}eobachtung und {R}eflexion. Vol.~1. Borntr{\"a}ger, Konigsberg.

\bibitem[{Von~Kupffer(1906)}]{von-kupffer1906.1-272}
Von~Kupffer, C., 1906. Die {Morphogenie des Zentralnervensystems}. In: Hertwig,
  O. (Ed.), {Handbuch der vergleichenden und experimentellen Entwicklungslehre
  der Wirbeltiere}. Vol. 2 (3). Fischer, Jena, pp. 1--272.

\bibitem[{Vulliemoz et~al.(2005)Vulliemoz, Raineteau, and
  Jabaudon}]{vulliemoz2005.87-99}
Vulliemoz, S., Raineteau, O., Jabaudon, D., 2005. Reaching beyond the midline:
  Why are human brains cross wired? Lancet Neurol. 4~(2), 87--99.

\bibitem[{Wicht and Northcutt(1990)}]{wicht1990.315-328}
Wicht, H., Northcutt, R.~G., 1990. Retinofugal and retinopetal projections in
  the pacific hagfish, {{\em Eptatretus stouti} (M}yxinoidea). Brain Behav.
  Evol. 36, 315--328.

\bibitem[{Wilson and Maden(2005)}]{wilson2005.1-13}
Wilson, L., Maden, M., 2005. The mechanisms of dorsoventral patterning in the
  vertebrate neural tube. Dev. Biol. 282~(1), 1--13.

\bibitem[{Zompa and Dubuc(1996)}]{zompa1996.221-227}
Zompa, I.~C., Dubuc, R., 1996. A mesencephalic relay for visual inputs to
  reticulospinal neurones in lampreys. Brain Res. 718~(1-2), 221--227.

\end{thebibliography}

\newpage

\appendix

\section{Additional Files}

The additional movies can be viewed and downloaded from \href{https://figshare.com/articles/media/An_embryological_twist_in_zebra_fish_embryo_as_revealed_by_cellular_movement_patterns_/24104085?file=42290463}{FigShare.com}, doi: \href{https://dx.doi.org/10.6084/m9.figshare.24104085}{10.6084/m9.figshare.24104085}.

\subsection{Additional movie 1: low.mov}\label{supp1}

Supplementary movie A.1 is based on the supplementary movie S2 of Keller et al. (2008), (with permission) $\copyright$ Keller \href{http://www.embl.de/digitalembryo/fish.html}{www.embl.de/digitalembryo}. For this we extracted a number of movie frames from their supplementary Movie S2. During this period (865-1000 min) the cells that will form the prospective eye cell masses, migrate clockwise (encircled blue and red respectively), whereas the future mid- and hindbrain rotate anti-clockwise (green arrows). The prospective right eye cell mass (blue) is initially invisible because it is hidden by the embryo. Starting from 865 min, every 15-min a frame was marked. For accurate playback, please select ``Play All Frames'' in the QuickTime Player. Note that the beginning and end of the movie are sped-up.

\subsection{Additional movie 2: selected.mov}\label{supp2}

 Movie A.2 is a sped-up version of Additional movie 1, to show more clearly the axial compensatory movements. Advice for repeated playing: select ``Loop'' in your Quicktime player.

\end{document}